\documentclass[conference]{IEEEtran}
\IEEEoverridecommandlockouts
\usepackage{stfloats} 
\usepackage{multirow} 
\usepackage{cite}
\usepackage{booktabs}
\usepackage{amsmath,amssymb,amsfonts}
\usepackage{amsmath}      
\usepackage{amssymb}      
\usepackage{algpseudocode}
\usepackage[ruled,norelsize,vlined,linesnumbered]{algorithm2e}
\usepackage{float}        
\makeatletter
\newcommand{\removelatexerror}{\let\@latex@error\@gobble}
\makeatother
\usepackage{graphicx}
\usepackage{textcomp}
\usepackage{xcolor}
\usepackage{subcaption}

\usepackage{geometry}
\def\BibTeX{{\rm B\kern-.05em{\sc i\kern-.025em b}\kern-.08em
    T\kern-.1667em\lower.7ex\hbox{E}\kern-.125emX}}
\begin{document}

\title{Adaptive Sampling and Joint Semantic-Channel Coding under Dynamic Channel Environment}

\author{
\IEEEauthorblockN{Zhiyuan Qi$^{1,2}$, Yulong Feng$^{4,5}$, Zhijin Qin$^{1,2,3}$}
\IEEEauthorblockA{$^{1}$Department of Electronic Engineering, Tsinghua University, Beijing, 100084, China \\
$^{2}$State Key Laboratory of Space Network and 
Communications, Beijing, 100084, China \\
$^{3}$Beijing National Research 
Center for Information Science and Technology, Beijing, 100084, China \\
$^{4}$State Key Laboratory of Mobile Network and Mobile Multimedia Technology, Shenzhen, 518055, China \\
$^{5}$ZTE Corporation, Nanshan District, Shenzhen, 518055, China \\
E-mail: qzy24@mails.tsinghua.edu.cn, qinzhijin@tsinghua.edu.cn}

 \thanks{
 This work is supported in part by the National Key Research and Development 
 Program of China under Grant 2023YFB2904300, and in part by the National 
 Natural Science Foundation of China (NSFC) under Grant 62293484.

 }
 }
\maketitle

\begin{abstract}
Deep learning enabled semantic communications are attracting extensive attention. However, most works normally ignore the data acquisition process and suffer from robustness issues under dynamic channel environment. In this paper, we propose an adaptive joint sampling-semantic-channel coding (Adaptive-JSSCC) framework. Specifically, we propose a semantic-aware sampling and reconstruction method to optimize the number of samples dynamically for each region of the images. According to semantic significance, we optimize sampling matrices for each region of the most individually and obtain a semantic sampling ratio distribution map shared with the receiver. Through the guidance of the map, high-quality reconstruction is achieved. Meanwhile, attention-based channel adaptive module (ACAM) is designed to overcome the neural network model mismatch between the training and testing channel environment during sampling-reconstruction and encoding-decoding. To this end, signal-to-noise ratio (SNR) is employed as an extra parameter input to integrate and reorganize intermediate characteristics. Simulation results show that the proposed Adaptive-JSSCC effectively reduces the amount of data acquisition without degrading the reconstruction performance in comparison to the state-of-the-art, and it is highly adaptable and adjustable to dynamic channel environments.
\end{abstract}

\IEEEpeerreviewmaketitle

\section{Introduction}
With the rapid development of communication technology and continuous expansion of communication scenarios in the future, the explosive increase in the magnitude of wireless data, as well as the heavy overhead of data acquisition devices, put forward higher demands on the redesign of communication architecture. Compared with traditional methods, semantic communications based on artificial intelligence (AI) focus on the purpose of transmitting meanings, instead of precisely recovering all original bits \cite{ref14}, and therefore improves spectral efficiency, solving the impact of performance deterioration caused by cliff effect in channels, which in turn reduces end-to-end network delay.    
                           
The power of deep learning (DL) and neural networks enabled semantic communications rooted in strong abilities of semantic understanding and representation to obtain relevant compressed symbols, allowing downstream task execution. As pointed out at \cite{ref9}, \cite{ref13}, Qin \textit{et al.} introduced essential theories of semantic communication. In recent years, research on semantic communication mainly paid attention to the design of encoder-decoder architecture and feature-extracting algorithms. Among them the most classical research was deep joint source-channel coding (DeepJSCC)\cite{ref0}, through employing deep neural networks to complete source and channel coding with a joint module at the same time, DeepJSCC significantly overcame the situation where it was almost impossible to transmit images by employing independent coding at low SNRs.

However, the procedure of data acquisition, which generated the source, was usually ignored in DeepJSCC-based systems, relatively independent from communication parts. As a beginning, the lack of sampling for semantic communication urged to be solved with joint design with channels. For semantic sampling a content-aware sensing method was proposed in the direction of compression sensing \cite{ref1}, completing fine-tuning of sensing based on a uniform matrix and semantic matrix. Song \textit{et al.} proposed a spatial adaptive sampling method, which captures and enhances the spatial dependency in the quantified latent representations based on a prior quality map and spatial importance, followed by probabilistic modeling to infer the quality mapping by optimizing the rate-task loss to maintain the image quality of the important regions \cite{ref2}. Inspired by that, Zhang \textit{et al.} improved the attentional mechanism for incorporating spatial intelligent perception \cite{ref3}. Another direction of downsampling was computational imaging, aiming at the intelligent design of camera code aperture for image for video content. Bacca in \cite{ref10} proposed an adjustable real-valued coded aperture in the depth optical imaging system according to semantics and specific task objectives to add the regular term penalty cost. Vargas \textit{et al.} in \cite{ref11} and Marwah \textit{et al.} in \cite{ref12} modeled the images captured by the camera as an angular projection of the incident light field, and designed a composite coding architecture of time-varying coded aperture and spatially coded shutter, which improved the quality of the imaging and the pixel stability both in time and space.

\begin{figure*}[tb]
	\centering
		\includegraphics[width=0.8\linewidth]{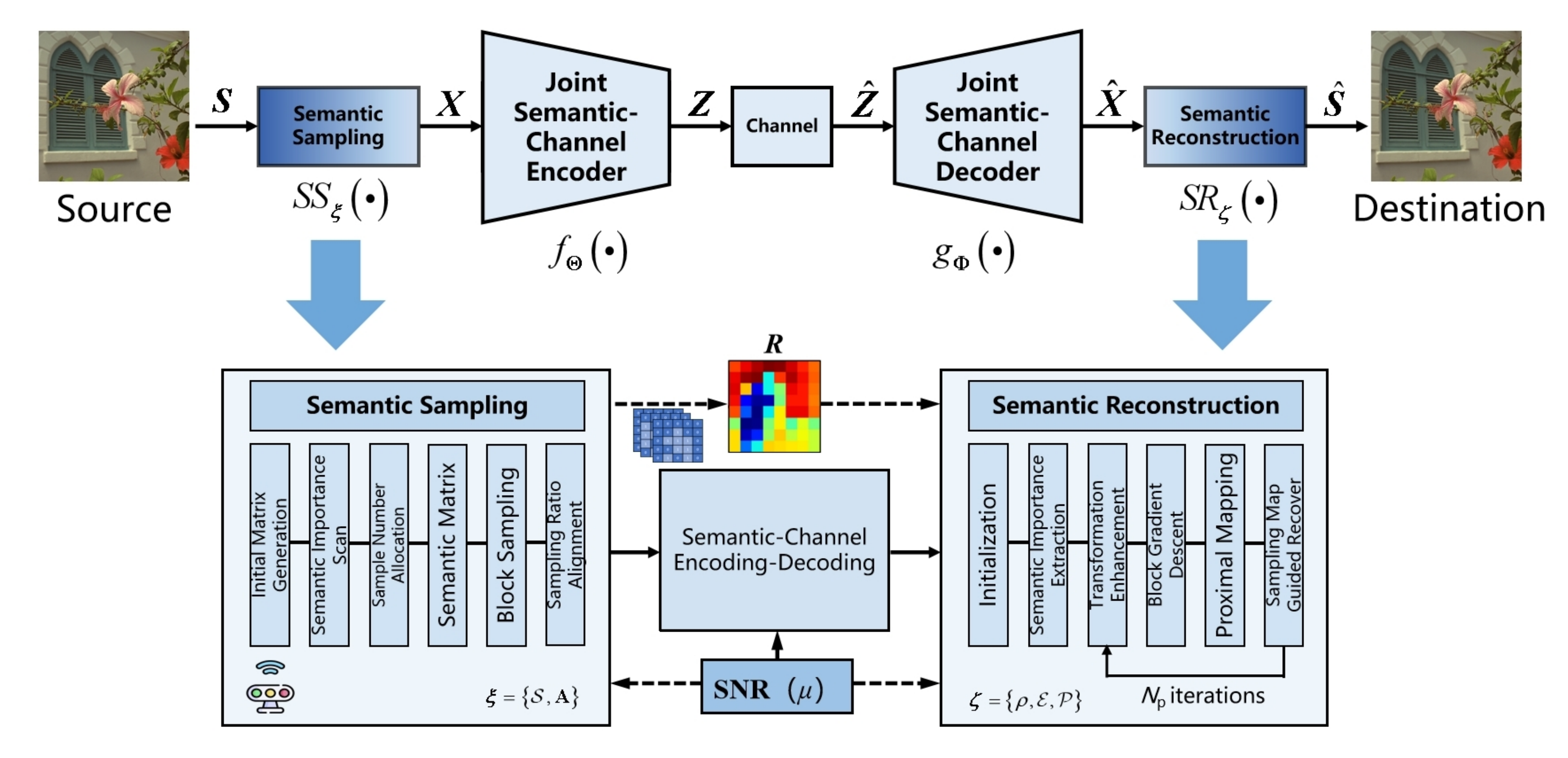}
	\caption{The framework of the proposed Adaptive-JSSCC, which contains semantic sampling, joint semantic-channel codec and reconstruction.}
	\label{fig-overall}
\end{figure*}

Moreover, given the fact that joint codec architecture for semantic communication based on a single SNR training cannot overcome the performance loss caused by the mismatch between the training and testing channel environment, and it was difficult to support image transmission in environments with variable SNR. Xu \textit{et al.} in \cite{ref4} proposed a method which is inputting SNR into neural network to scale original features. Inspired by \cite{ref4}, another method with respect to compression was proposed in which refined the adaptation to channel bandwidth ratio (CBR) to adjust the various compression of source \cite{ref5}. To improve system robustness to different modulation modes, Tung \text{et al.} proposed DeepJSCC-Q which satisfied constellations and hardware communication standard matching employing adaptive variational auto encoder architecture \cite{ref15}.

In this work, the main contributions of this paper are summarized as follows:

\begin{itemize}
    \item An adaptive joint sampling-semantic-channel coding framework under dynamic channel environment named Adaptive-JSSCC is proposed to effectively enhance the performance of transmitting images. 
    \item An attention-based channel adaptive module (ACAM) is proposed to further utilize characteristics of source and channel. The key designs of the novel codec architecture and semantic sampling and reconstruction algorithms are elaborated from the module.

\end{itemize}


    \section{System model}
In this section, we provide an overview about the system model of Adaptive-JSSCC as shown in Fig.~\ref{fig-overall}, including the components and structure of networks, introducing the processing flow briefly. We focus on reconstruction-oriented image transmission tasks considering channels and noise.

Semantic sampling module $S S_{\boldsymbol{\xi}}(\cdot)$, encoder $f_{\Theta}(\cdot)$, decoder $g_{\Phi}(\cdot)$, reconstruction module $S R_{\boldsymbol{\zeta}}(\cdot)$ are the sections where the corresponding parameter sets $\boldsymbol{\xi}$, $\Theta$, $\Phi$, $\boldsymbol{\zeta}$ can be trained end-to-end, while the channel cannot. Let the images to be transmitted are $\boldsymbol{s}\in\mathbb{R}^n$, where $n=H\times W\times C$ is determined as the dimension of original image, and $H,W,C$ respectively correspond to the height, width, channel numbers of an image. 

Firstly, at the transmitter, the original image $\boldsymbol{s}$, overall sampling ratio $r$ and SNR $\mu$ are inputted into $S S_{\boldsymbol{\xi}}(\cdot)$ to refine key data $\boldsymbol{x}\in\mathbb{R}^n$ from source. Meanwhile, a significant semantic sampling ratio distribution map $\boldsymbol{R}$ is obtained, which is shared with the receiver. Then, encoding function realizes the mapping $f_{\Theta}:\mathbb{R}^n\times\mathbb{R}\rightarrow\mathbb{C}^k$ to embed it into specific feature space with the help of SNR, where the encoded symbols $\boldsymbol{z}\in\mathbb{C}^k$ are the complex-value symbols. $k$ denotes the channel dimension, as well as the shape of output features. Generally, we have $k<n$, where $k$ represents channel bandwidth and $n$ represents source bandwidth. Thus, we refer to $k/n$ as the channel bandwidth ratio (CBR), which indicates the degree of the compressed source. The procedure of sampling and encoding can be respectively described as (\ref{eq-2-1}) and (\ref{eq-2-2}):
\begin{equation}
    \label{eq-2-1}
    \boldsymbol{x}=S S_{\boldsymbol{\xi}}(\boldsymbol{s},\mu,r),
\end{equation}
\begin{equation}
    \label{eq-2-2}
    \boldsymbol{z}=f_{\Theta}(\boldsymbol{x},\mu).
\end{equation}



Secondly, the encoded symbols are transmitted through Additive White Gaussian Noise (AWGN) denoted by $\eta(\cdot)$. The channel can be modeled as (\ref{eq-2-5}), where $\boldsymbol{n} \sim \mathcal{C N}\left(0, \sigma^2 \mathbf{I}\right)$ represents noise:
\begin{equation}
    \label{eq-2-5}
    \hat{\boldsymbol{z}}=\eta(\boldsymbol{z})=\boldsymbol{z}+\boldsymbol{n}.
\end{equation}

Thirdly, at the receiving end, the decoder addresses the transmitted symbol $\hat{\boldsymbol{z}}$ and SNR $\mu$, realizing the mapping $g_{\Phi}:\mathbb{C}^k\times\mathbb{R}\rightarrow\mathbb{R}^n$, where ${\hat{\boldsymbol{x}}\in\mathbb{R}^n}$ are the decoding results. According to $\hat{\boldsymbol{x}}$, $\mu$ and semantic distribution map $\boldsymbol{R}$, we implement semantic reconstruction. The above two processes can be described as:
\begin{equation}
    \label{eq-2-6}
    \hat{\boldsymbol{x}}=g_{\Phi}(\hat{\boldsymbol{z}}, \mu)=g_{\Phi}\left(\eta\left(f_{{\Theta}}(\boldsymbol{x}, \mu)\right), \mu\right),
\end{equation}
\begin{equation}
    \label{eq-2-7}
    \hat{\boldsymbol{s}}=S R_{\boldsymbol{\zeta}}(\hat{\boldsymbol{x}}, \mu, \boldsymbol{R}).
\end{equation}

Finally, based on the distortion between the original image and the final transmitted reconstructed image $\boldsymbol{d}(\boldsymbol{s}, \hat{\boldsymbol{s}})=\frac{1}{n} \sum_{i=1}^n\left(\boldsymbol{s}_i-\hat{\boldsymbol{s}}_i\right)^2$, the optimization objective of the end-to-end joint design is to obtain the optimal set of parameters $\boldsymbol{\xi}^*, \Theta^*, \Phi^*, \boldsymbol{\zeta}^*$ that minimizes the desired distortion under a certain CBR, and the details for these parameters are shown in Fig.\ref{fig-overall}. To simplify the representation, we set $\boldsymbol{\Theta}$ include all the parameters involved in Adaptive-JSSCC:
\begin{equation}
    \label{eq-2-8}
    \left(\boldsymbol{\xi}^*, \Theta^*, \Phi^*, \boldsymbol{\zeta}^*\right)=\underset{\boldsymbol{\xi}, \Theta, \Phi, \boldsymbol{\zeta}}{\arg \min } \mathbb{E}_{p(r)} \mathbb{E}_{p(\mu)} \mathbb{E}_{p(\boldsymbol{s},\hat{\boldsymbol{s}})}[\boldsymbol{d}(\boldsymbol{s}, \hat{\boldsymbol{s}})].
\end{equation}

\section{Proposed System Components}
In this section, we introduce a module to study the condition of the channel environment given SNRs. Employing the module we provide a detailed description of the adaptive sampling and reconstruction algorithms for joint semantic and channel characterization, being applied to a new designed encoder-decoder architecture of semantic communication.

\subsection{Attention-based Channel Adaptive Module}
To scale sampling results or representation of encoded and decoded symbols according to dynamic communication quality, we propose an attention-based channel adaptive module (ACAM) to capture the relationship between input features and channel, generating different scaling parameters for channel characteristics through the channel soft attention \cite{ref4}, so as to maintain high quality during sampling-reconstruction and encoding-decoding under different SNRs.

Specifically, let the features of the input module be $F$. ACAM first performs global average pooling along each channel on $F$ to obtain $I$. Then we concatenate $I$ with SNR of the current channel environment to obtain the intermediate feature $I^{\prime}$ by:
\begin{equation}
    \label{eq-2-9}
    I^{\prime}=\operatorname{concat}(\mu,I), 
\end{equation}
where $\operatorname{concat}(F_1,F_2)$ denotes the splicing of feature $F_1$ and feature $F_2$. Next, $I^{\prime}$ is inputted into a nonlinear deep neural network to generate the scaled factor $\kappa$:
\begin{equation}
    \label{eq-2-10}
    \kappa=\boldsymbol{\sigma_2}(\boldsymbol{w_2}(\boldsymbol{\sigma_1}(\boldsymbol{w_1}I^{\prime}+\boldsymbol{b_1})+\boldsymbol{b_2}),
\end{equation}
where $\boldsymbol{\sigma_1}$, $\boldsymbol{\sigma_2}$ denote PReLU layer and Sigmoid layer respectively, and $\boldsymbol{w_1}$, $\boldsymbol{w_2}$, $\boldsymbol{b_1}$, $\boldsymbol{b_2}$ are parameters of linear layers. Finally, an element-to-element product $\odot$ of $\kappa$ and $F$ is performed, which is tuned for each element of the input features and then obtain the channel environment adaptive features $F^{\prime}$:
\begin{equation}
    \label{eq-2-11}
    F^{\prime}=\kappa \odot F.
\end{equation}

\begin{figure}[!t]
	\centering
		\includegraphics[width=1.0\linewidth]{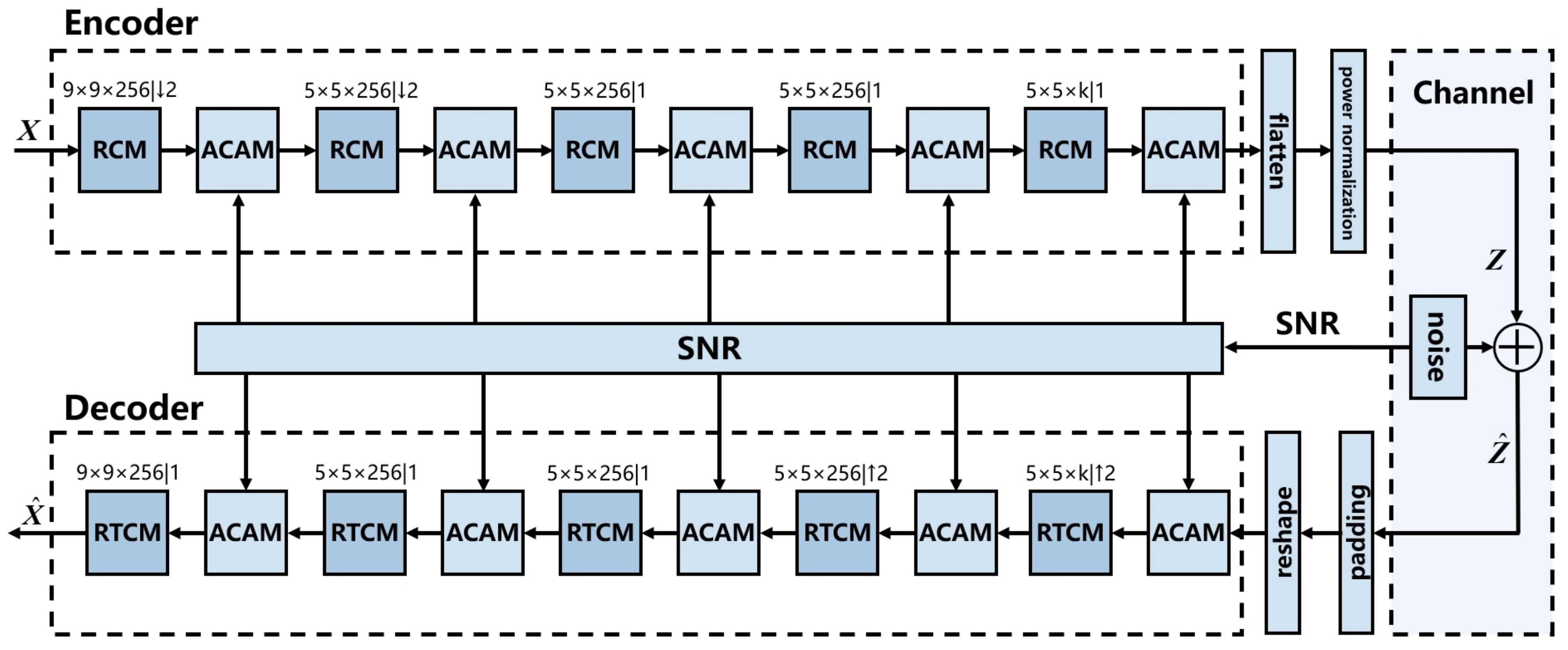}
	\caption{The proposed joint semantic-channel codec.}
	\label{fig-codec}
\end{figure}

As shown in Fig.~\ref{fig-overall} and Fig.~\ref{fig-codec}, ACAM appears in semantic sampling and reconstruction, encoder and decoder when there is a need for adapting to the current channel environment, studying the exacter representation of the transmitted symbol while processing. First of all, we discuss the semantic encoder and decoder. Based on classical DeepJSCC systems, we design a new encoder-decoder architecture in Fig.~\ref{fig-codec}, enabling adaptive adjustment according to SNRs in the current environment during encoding and decoding. Residual convolutional module (RCM) and residual transposed convolutional module (RTCM) employ residual concatenation to improve the original convolutional block, where $M\times N\times C\mid S\downarrow\uparrow$ denotes shape and output channels of kernels, the symbols $\downarrow$ and $\uparrow$ denote down-sampling and up-sampling respectively, and the parameter $S$ denotes the step size. Our design enhances the semantic-channel joint encoding and decoding.

\subsection{Semantic Sampling}
In the proposed image transmission system, we design a semantically based down-sampling method that is adaptive to dynamic channel environment. Given a preset overall sampling ratio $r$, according to a specific strategy, we pay more attention to the regions with more significant semantic information in which we assign samples of specific numbers. 
The concrete sampling procedure is shown in Algorithm 1.

\begin{algorithm}[!t]
\caption{Adaptive Semantic Sampling Algorithm}
      \KwIn{Overall Sampling Ratio $r$, Original Images $\boldsymbol{S}$, Initial Sampling Matrix $\boldsymbol{{\rm A}}$ generated by SVD, SNR $\mu$, Semantic Scanning Network $\mathcal{S}$.}
      \KwOut{Semantic Sampling Ratio Distribution Map $\boldsymbol{R}$, Sampled Symbols $\boldsymbol{X}$.}
    \BlankLine
    Divide $\boldsymbol{S}$ into ${\{\boldsymbol{s_i}\}}^{l}_{i=1}$\;
    $\boldsymbol{M} \leftarrow \mathcal{S}, \boldsymbol{S}, \mu$\;
    \For{$i$ in $l$ s.t. overall sampling ratio $r$}{
    Obtain specific sampling ratio for each block: $r_i \leftarrow r, \boldsymbol{M}$\;
    Obtain semantic sampling matrix for each block: $\boldsymbol{{\rm {A}}_{q_i}} \leftarrow \boldsymbol{{\rm A}}, r_i$\;
    Sampling: $\boldsymbol{y_i}=\boldsymbol{{\rm {A}}_{q_i}}\boldsymbol{s_i}$\;
    Aligning: $\boldsymbol{x_i}=\boldsymbol{{\rm {A}}^{\rm{T}}_{q_i}}\boldsymbol{y_i}$\;
}
    Transform and revise: $\boldsymbol{R} \leftarrow {\{r_i\}}^{l}_{i=1}, \boldsymbol{M}$\;
    \textbf{return} $\boldsymbol{R}, \boldsymbol{X}={\{\boldsymbol{x_i}\}}^{l}_{i=1}$;
\end{algorithm}
Firstly, we divided $S$ into non-overlapping blocks $\left\{\boldsymbol{s}_i\right\}_{i=1}^l$ of size $N=B\times B$, where the edge length of each single square block is denoted as $B$. Since the image size is $H\times W$, the number of blocks is $l=(H/B)\times(W/B)$. When sampling is performed, each block corresponds to a sampling matrix. That is, our semantic sampling is block by block.

\begin{figure}[!t]
	\centering
		\includegraphics[width=0.65\linewidth]{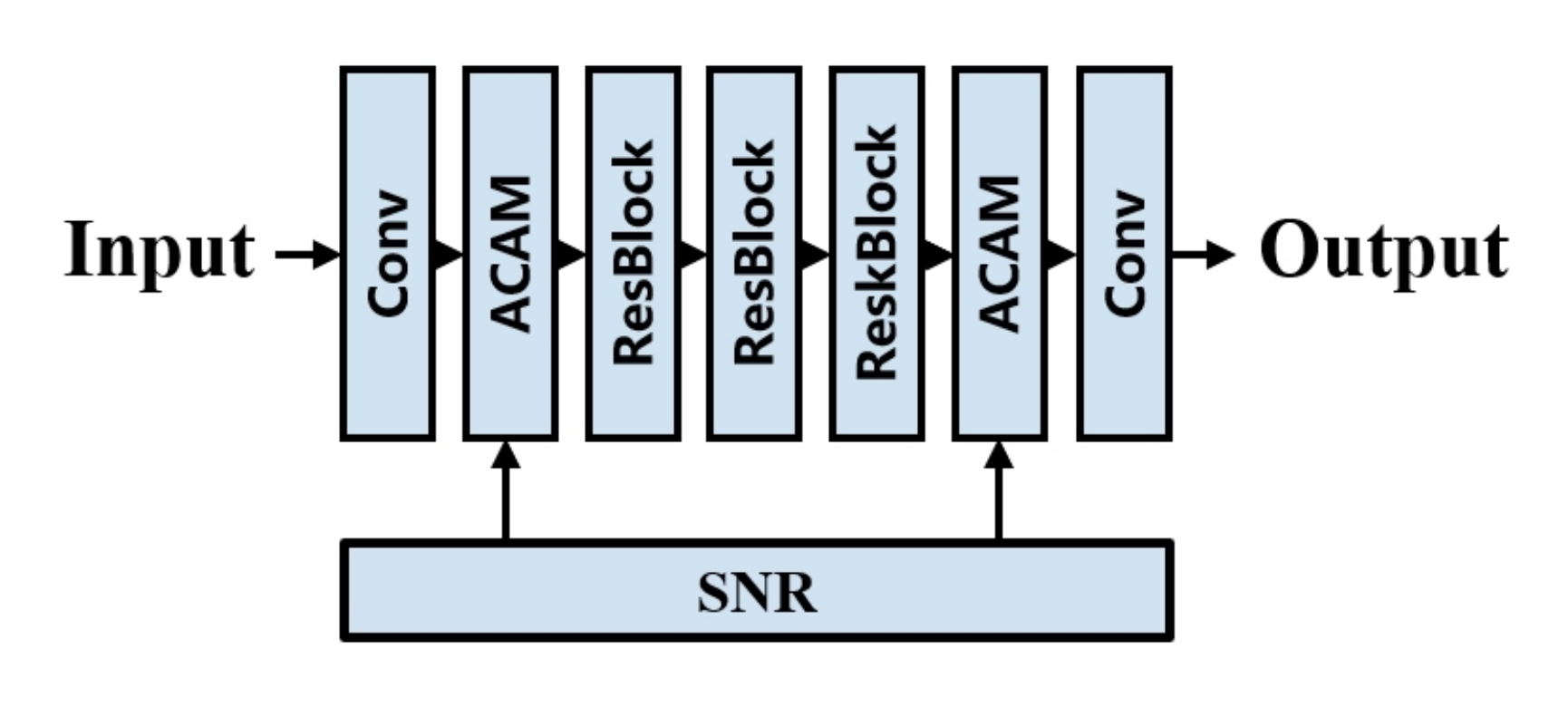}
	\caption{The architecture of semantic scanning network.}
	\label{fig-scan}
\end{figure}

To evaluate the saliency of semantic information at each block and highlight the importance of different regions, we design a channel environment adaptive semantic scanning network $\mathcal{S}$ as shown in Fig.~\ref{fig-scan}. Through inputting initial images $S$ and SNR $\mu$ we can get saliency distribution map $\boldsymbol{M}=\mathcal{S}(\boldsymbol{S},\mu)$. Then, we employ overall sampling ratio $r$ and $\boldsymbol{M}$ to address initial matrix $\boldsymbol{{\rm A}}$ and then obtain semantic sampling matrix $\boldsymbol{{\rm {A}}_{q_i}}$, as well as individual sampling ratio $r_i=q_i/N$ for each block. For the sake of realizing the precise assignment of the sampling ratio of each region in the image, we implement aggregation on $\boldsymbol{M}$ to obtain semantic sampling ratio distribution map $\boldsymbol{R}$ while sharing it with the receiver. Here we consider a situation in which the transmission of $\boldsymbol{R}$ is without any loss. The device works, which conducts compressed sensing for corresponding blocks, obtains initial sampled results ${\{\boldsymbol{y_i}\}}^{l}_{i=1}$, and then aligns the sampling ratios block by block for subsequent semantic encoding. Finally we obtain suitable symbols ${\{\boldsymbol{x_i}\}}^{l}_{i=1}$ for transmitting.



\subsection{Semantic Reconstruction}
This part provides a detailed description of semantic reconstruction method that perfectly matches sampling, mainly consisting of proximal gradient descent (PGD) optimization and proximal mapping\cite{ref7}. The procedure of reconstruction is guided by the semantic distribution map $\boldsymbol{R}$ shared with the transmitter. To parse semantic information in $\boldsymbol{R}$, the semantic extracting network $mathcal{E}$ possesses the same structure shown in Fig.~\ref{fig-scan}, and then drives the proximal mapping network to complete high-quality image reconstruction after $N_p$ rounds of iterations. The objective of the whole reconstruction optimization process in each iteration is:
\begin{equation}
    \label{eq-2-14}
\hat{\boldsymbol{x}}_i^{(k)}=\underset{\boldsymbol{x}_i}{\arg \min } \frac{1}{2}\left\|\boldsymbol{x}_i-\boldsymbol{\hat{v}}_i^{(k)}\right\|_2^2+\lambda \mathcal{R}\left(\boldsymbol{x}_i\right) ,
\end{equation}
where $\lambda\mathcal{R}(\boldsymbol{x}_i)$ is regularization term.

\begin{algorithm}[!t]
\caption{Adaptive Semantic Reconstruction Algorithm}
      \KwIn{Decoder Output Symbols $\hat{\boldsymbol{X}}$, Semantic Sampling Ratio Distribution Map $\boldsymbol{R}$, SNR $\mu$, Semantic Extracting Network $\mathcal{E}$, Gradient Descent Step $\rho$, Proximal Mapping Network $\mathcal{P}$,  Iteration Rounds $N_p$.}
      \KwOut{Reconstructed Images $\hat{\boldsymbol{S}}$.}
    \BlankLine
    \textbf{Initialization and enhancement:} $\boldsymbol{\hat{X}}^{(0)} \leftarrow \boldsymbol{\hat{X}}$\; 
    \For{$k$ in $N_p$}{
    $\boldsymbol{M}^{\prime} \leftarrow \mathcal{E}^{(k)}, \boldsymbol{R}, \mu$\;
    Unfold into blocks: ${\{\hat{\boldsymbol{x}}^{(k)}_i\}}^{l}_{i=1} \leftarrow \boldsymbol{\hat{X}}^{(k)}$\;
    Gradient descent by blocks: ${\{\hat{\boldsymbol{v}}^{(k)}_i\}}^{l}_{i=1} \leftarrow {\{\hat{\boldsymbol{x}}^{(k)}_i\}}^{l}_{i=1}, \mathcal{\rho}^{(k)}$\;
    Concatenate features and reconstruct: ${\{\hat{\boldsymbol{s}}^{(k)}_i\}}^{l}_{i=1} \leftarrow {\{\hat{\boldsymbol{v}}^{(k)}_i\}}^{l}_{i=1},\boldsymbol{M}^{\prime}, \mathcal{P}^{(k)}$\;
    Fold and update: $\boldsymbol{\hat{X}}^{(k+1)} \leftarrow \boldsymbol{\hat{S}}^{(k)}$\;}
    \textbf{return} $\hat{\boldsymbol{S}}$;
\end{algorithm}

As shown in Algorithm 2, firstly, we perform random data enhancement to decoder output symbols at the receiver end, and then in k-th iteration each block is allowed to complete the update of the proximal gradient to obtain the intermediate result $\boldsymbol{\hat{v}}_i^{(k)}$. Then, the sampling map $\boldsymbol{R}$ shared with the transmitter is processed and fed into the channel environment adaptive semantic extraction network to obtain the feature map $\boldsymbol{M^{\prime}}=\mathcal{E}^{(k)}\left(\boldsymbol{R}, \mu\right)$, and $\boldsymbol{v}_i^{(k)}$ is spliced with $\boldsymbol{M^{\prime}}$ and fed into the proximal mapping network to obtain the reconstruction results in a single round:
\begin{equation}
    \label{eq-2-13}
\boldsymbol{\hat{s}}_i^{(k)}=\boldsymbol{\hat{v}}_i^{(k)}+\mathcal{P}^{(k)}\left(\operatorname{concat}\left(\boldsymbol{\hat{v}}_i^{(k)}, \mathcal{E}^{(k)}\left(\boldsymbol{R}, \mu\right)\right)\right) ,
\end{equation}
where $\mathcal{P}^{(k)}$ denotes deep proximal mapping network. As the separate processing based on chunking makes the edge effect of different chunks obvious, the splicing will produce more obvious artifacts. $\mathcal{P}^{(k)}$ is capable of fully perceiving the relationship between blocks and effectively dealing with the extra noise caused by the traces of splicing. The last turn outputs $\{\boldsymbol{\hat{s}}_i^{(N_p)}\}_{i=1}^{l}$ represent the final reconstruction results.

\section{Simulation Results}
In this section, we conduct ablation experiments and the simulation results demonstrate the performance of Adaptive-JSSCC in various aspects.
\subsection{Implementation Details}
\paragraph{Dataset and Metrics} 
The training dataset with the number of images is $N_b=25600$, randomly cropped and rotated image blocks of size 128$\times$128 from super-resolution dataset T91\cite{ref7} and another dataset Train400\cite{ref8}. 

In order to measure the quality of transmitted images compared with ground truth, we choose peak signal-to-noise ratio (PSNR) and structural
similarity index (SSIM) as the evaluation metrics for the downstream task of image reconstruction.

\paragraph{Training Settings}
We train Adaptive-JSSCC model end-to-end where the block size $B=32$ for sampling and reconstruction and therefore the number of blocks is $l=16$. The codec architecture and algorithms are implemented based on PyTorch. The optimization iteration rounds for semantic reconstruction is $N_p=11$. Train SNR is uniformly distributed in the range of $[0, 20]$ dB. In addition, for the overall sampling ratio $r=q/N$, where $q$ is randomly equivocally selected in the range of $[1, N]$. A total of 300 epochs of training are conducted with the learning rate of the first 200 epochs being $10^{-4}$, the learning rates of from 200-th to 260-th epoch being $10^{-5}$. The last 40 epochs are fine-tuning with a learning rate of $10^{-6}$. Batch size in training is 16. The optimizer is the Adam optimizer. The loss function of end-to-end optimization is defined as follows:
\begin{equation}
    \label{eq-3-1}
\mathcal{L}_{\boldsymbol{\Pi}}=\frac{1}{l N N_b} \sum_{j=1}^{N_b}\left\|\mathcal{F}_{\text {Adaptive-JSSCC}}\left(\boldsymbol{S}_j, q_j, \mu ; \boldsymbol{\Pi}\right)-\boldsymbol{S}_j\right\|_2^2 ,
\end{equation}
where $\boldsymbol{\Pi}=\{\mathcal{S}, \mathbf{A}\} \cup \Theta \cup \Phi \cup\left\{\rho, \mathcal{E}, \mathcal{P}\right\}$ is the set of all trainable parameters, and $\mathcal{F}_{\text {Adaptive-JSSCC}}(\cdot)$ denotes the function of end-to-end transmitting through Adaptive-JSSCC.

\paragraph{Ablation Study Settings}

\begin{figure}[!t]
	\centering
		\includegraphics[width=0.9\linewidth]{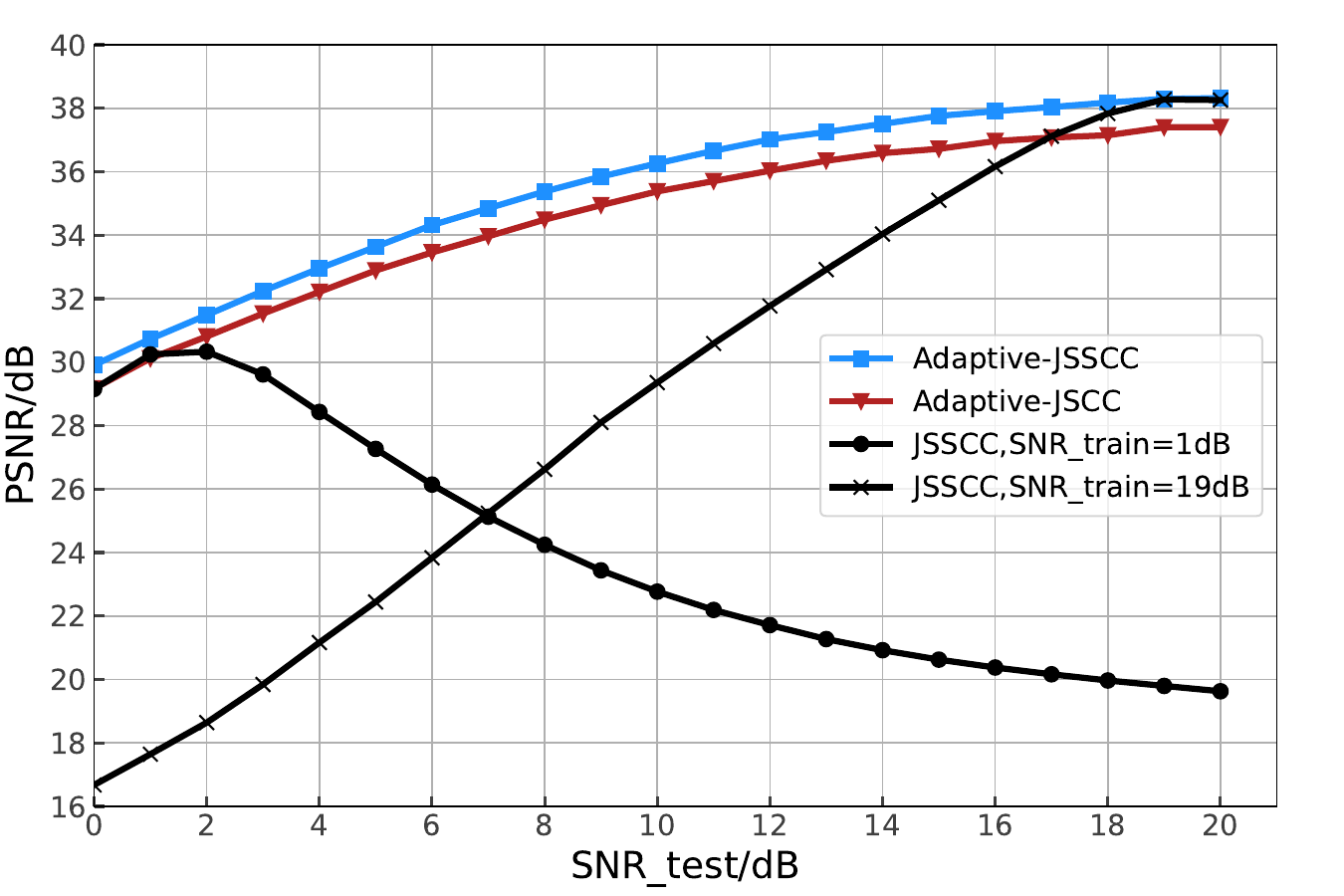}
	\caption{Comparison of model performance at different SNRs.}
	\label{fig-mismatch}
\end{figure}

\begin{table}[!t]
\caption{Model ablation implementation details.}
\begin{center}
\begin{tabular}{c|c|c|c}
\hline
 & \textbf{Semantic/Uniform}& & \\
\textbf{Models}&\textbf{Sampling}&\textbf{ACAM(w/o)}&\textbf{SNR\underline{ }train}\\
 &\textbf{-Reconstruction(w/o)}& & \\
\hline
Adaptive-JSSCC& \checkmark & \checkmark &[0,20] dB \\
Adaptive-JSCC& & \checkmark & [0,20] dB \\
JSSCC& \checkmark & & 1 dB,19 dB\\
JSCC& & & 1 dB,19 dB \\
\hline
\end{tabular}
\label{tab-1}
\end{center}
\end{table}




To verify the functionality and validity under dynamic channel environment, and show the performance as the sampling ratio decreases, except for Adaptive-JSSCC, we design several sets of ablation models called Adaptive-JSCC, JSSCC and JSCC where the constitutive components of each model are shown in Table~\ref{tab-1}.

\subsection{Performance of Adaptive-JSSCC}
\paragraph{Performance Verification of Channel Environment Adaptation}
Fig.~\ref{fig-mismatch} indicates that for JSSCC without ACAMs, the channel mismatch between training and testing SNRs will lead to the degradation of reconstructed image quality. Whereas, the performance of our proposed Adaptive-JSSCC performs more gently and slowly under the same situations. In addition, compared to uniform sampling-reconstruction, the semantic method has an overall performance improvement of about 1dB at arbitrary SNR of any channel environment.

Table~\ref{tab-2} explores the optimization of the Adaptive-JSSCC on the storage overhead while $r$ is defined as 0.50 evaluated on Kodak24\cite{ref18}. To obtain better performance over a wide range of testing SNRs, it is necessary to store multiple JSSCC models training at different SNRs. JSSCC-n denotes that there are $n$ JSSCC models in one node of a network. Adaptive-JSSCC only occupies one model-size storage and performs better than an aggregation of multiple models. PSNR in Table~\ref{tab-2} is calculated as a better average PSNR of models' performance at all SNRs.

\paragraph{Performance Validation of Semantic Sampling-Reconstruction} 

Due to the objective of reducing data acquisition, the semantic sampling algorithm we proposed achieves a lower overall sampling ratio $r$ at the same SNR. That is, Adaptive-JSSCC completes the transmission with fewer samples under the same communication quality requirements, which in turn reduces the cost of device imaging and consumption of energy. Table~\ref{tab-3} indicates evaluation results of semantic and uniform methods at different overall sampling ratios. The semantic method typically yields an average PSNR improvement of the entire testing set by around 0.5 to 1 dB.

\begin{figure}[!t]
	\centering
		\includegraphics[width=1.1
  \linewidth]{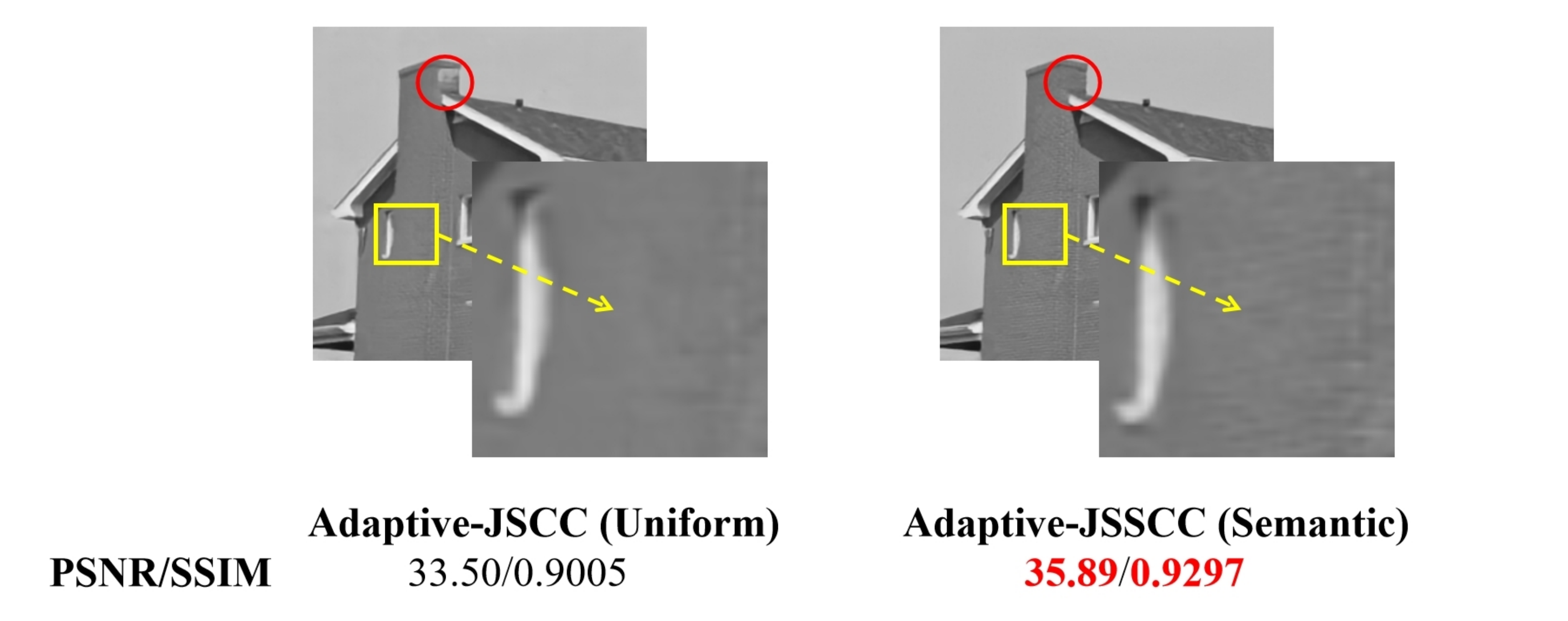}
	\caption{Details with uniform (left) and semantic (right) method of sampling-reconstruction.}
	\label{fig-compare}
\end{figure}

\begin{table}[!t]
\caption{Memory consumption of models.}
\begin{center}
\begin{tabular}{c|c|c}
\hline
\textbf{Models} & \textbf{Memory Consumption}& \textbf{PSNR}\\
\hline
Adaptive-JSSCC & 102.58MB & \textbf{35.45dB} \\
JSSCC& \textbf{102.53MB} & 28.65dB \\
JSSCC-2& 205.06MB & 31.55dB\\
 JSSCC-5& 512.66MB & 34.82dB\\
\hline
\end{tabular}
\label{tab-2}
\end{center}
\end{table}



As shown in Fig.~\ref{fig-compare}, we observe and compare the performance of semantic-based and semantic-free sampling and reconstruction at a sampling ratio of 0.50 and a SNR of 3 dB. Higher quality of restoration as well as more complete details can be obtained by guiding the image with semantic information, recovering more details of edges and textures in the image.

\begin{figure}[!t]
	\centering
	\begin{subfigure}{1\linewidth}
		\centering
		\includegraphics[width=0.67\linewidth]{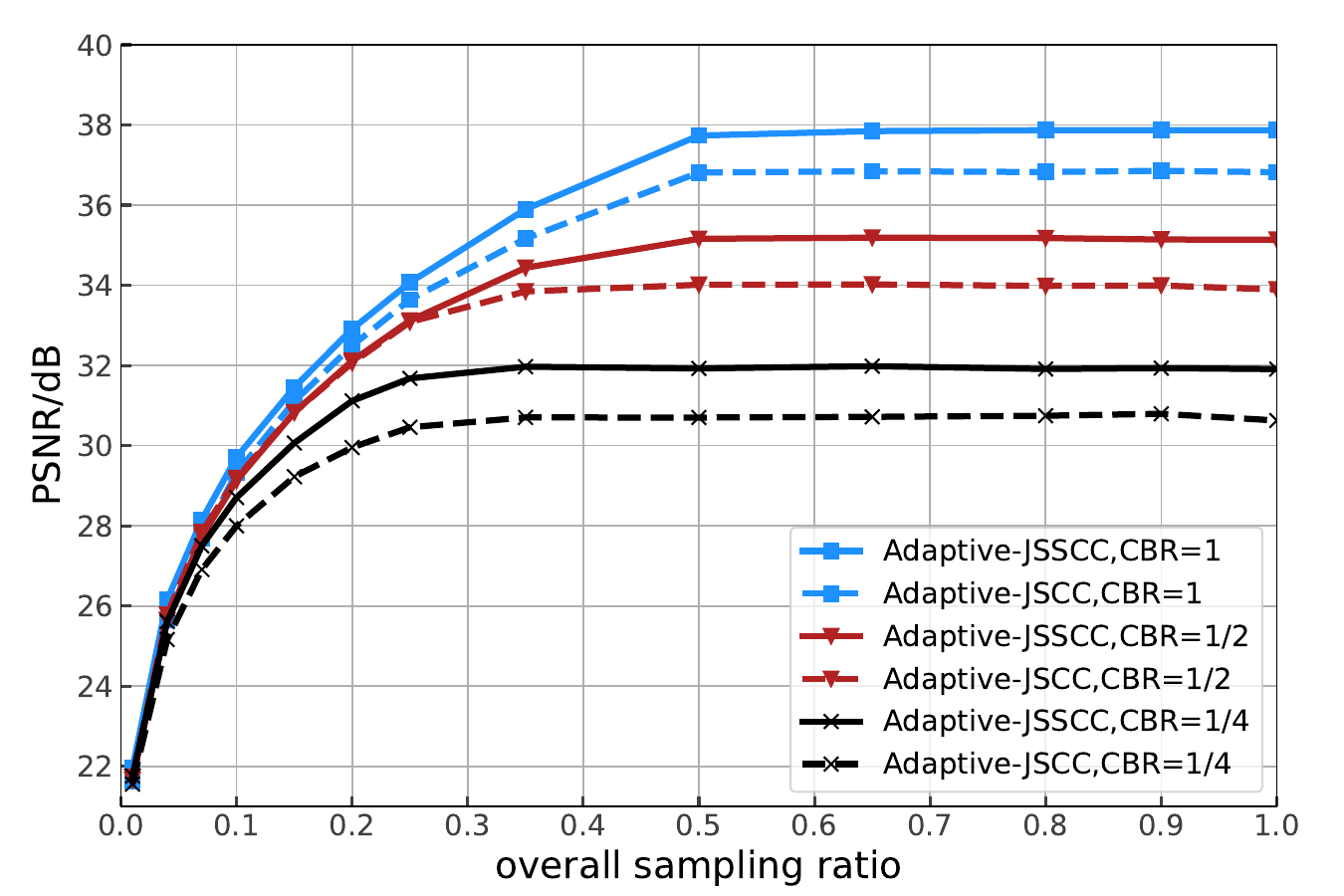}
		\caption{PSNR performance varies in CBRs}
		\label{6a}
	\end{subfigure}
	\centering
	\begin{subfigure}{1\linewidth}
		\centering
		\includegraphics[width=0.67\linewidth]{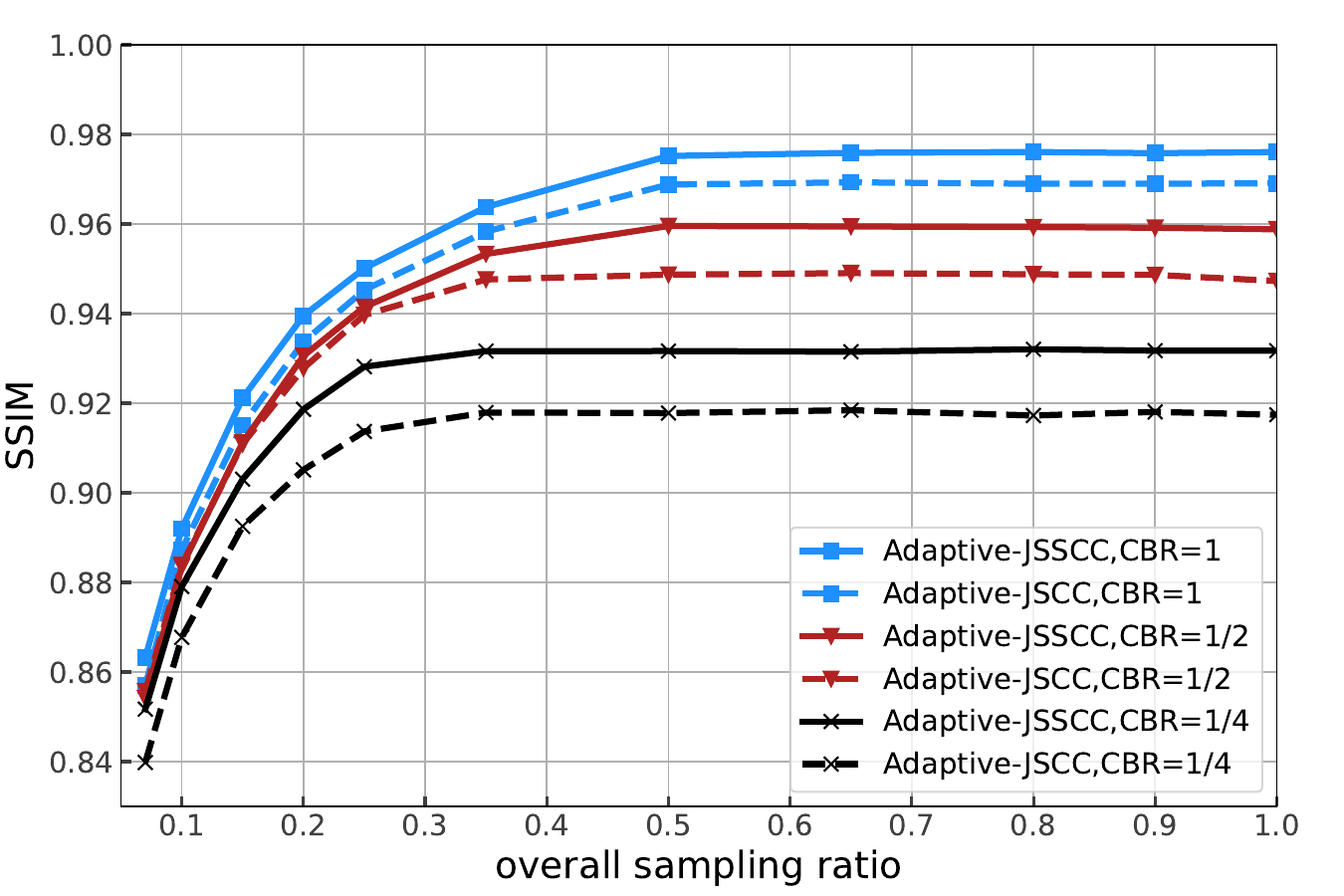}
		\caption{SSIM performance varies in CBRs}
		\label{6b}
	\end{subfigure}
	\centering
	\caption{
Performance comparison between semantic and uniform sampling-reconstruction under different CBRs when the sampling ratio varies.}
	\label{fig-two}
\end{figure}

Fig.~\ref{fig-two} compares Adaptive-JSSCC with uniform method under different compression while achieving the same SNR of 15dB, and an overall sampling ratio of 0.50. Channel bandwidth ratio (CBR) demonstrates the amount of remaining compression space after encoding. CBR deterioration brings poor PSNR and SSIM of the reconstructed image, representing the transmission quality. What's more, we figure out that performance achieves saturation earlier in the models trained at lower CBR. 

\begin{table*}[t]
\caption{Semantic and uniform sampling-reconstruction performance evaluated on testing sets Set11/Kodak24/BSD68.}
\begin{center}
\begin{tabular}{c|c|cccccc}
\hline
\multirow{2}{*}{\textbf{Testset}} & \multirow{2}{*}{\textbf{Method}} & \multicolumn{6}{c}{\textbf{PSNR(dB)/SSIM under Various Sampling Ratios \textit{r}}}     \\ \cline{3-8} 
                         &                         & $r=0.01$         & $r=0.04$         & $r=0.10$         & $r=0.30$         & $r=0.40$         & $r=0.50$         \\ \hline
\multirow{2}{*}{Set11\cite{ref17}}   & Adaptive-JSSCC                & 21.97/0.6062 & 26.16/0.8087 & 29.66/0.8920 & 35.03/0.9579 & 36.62/0.9687 & \textbf{37.73/0.9751} \\ \cline{2-8} 
                         & Adaptive-JSCC                 & 21.57/0.5831 & 26.65/0.7980 & 29.30/0.8860 & 34.49/0.9530 & 35.82/0.9627 & 36.80/0.9689 \\ \hline
\multirow{2}{*}{Kodak24\cite{ref18}} & Adaptive-JSSCC                & 23.73/0.5947 & 26.53/0.7329 & 29.19/0.8376 & 33.85/0.9377 & 35.48/0.9564 & \textbf{36.77/0.9677} \\ \cline{2-8} 
                         & Adaptive-JSCC                 & 23.60/0.5846 & 26.29/0.7231 & 28.85/0.8304 & 33.44/0.9324 & 34.91/0.9505 & 36.05/0.9612 \\ \hline
\multirow{2}{*}{BSD68\cite{ref19}}   & Adaptive-JSSCC                & 22.82/0.5499 & 25.53/0.7001 & 28.01/0.8142 & 32.29/0.9276 & 33.85/0.9497 & \textbf{35.07/0.9628} \\ \cline{2-8} 
                         & Adaptive-JSCC                 & 22.41/0.5398 & 25.22/0.6914 & 27.69/0.8076 & 31.88/0.9222 & 33.31/0.9440 & 34.42/0.9569 \\
\hline
\end{tabular}
\label{tab-3}
\end{center}
\end{table*}


\section{CONCLUSION}
In this work, we develop a semantic communication for image transmission, based on an adaptive joint sampling-semantic-channel coding scheme under dynamic channel environment, named Adaptive-JSSCC. The methods we propose not only consider reducing the expense of data acquisition but also overcome poor performance caused by channel mismatch between training and testing SNRs. Particularly, an attention-based channel adaptive module and adaptive semantic sampling-reconstruction algorithms are designed. Simulation results demonstrate that our methods significantly reduce the overhead of data acquisition while keeping high reconstruction quality under dynamic channel environment in terms of SNRs and CBRs.

\end{document}